\documentclass[aps,12pt,final,notitlepage,oneside,onecolumn,nobibnotes,nofootinbib,superscriptaddress,noshowpacs,centertags]{revtex4}

\begin{document}

\title{$\pi\pi$ scattering in a nonlocal Nambu -- Jona-Lasinio model}

\author{\firstname{A. A.} \surname{Osipov}}
\email{osipov@nu.jinr.ru}
\affiliation{Dzhelepov Laboratory of Nuclear Problems, \\
Joint Institute for Nuclear Research, 141980 Dubna, Russia}
\author{\firstname{A. E.} \surname{Radzhabov}}
\email{aradzh@theor.jinr.ru}
\affiliation{Bogoliubov Laboratory of Theoretical Physics, \\
Joint Institute for Nuclear Research, 141980 Dubna, Russia}
\author{\firstname{M. K.} \surname{Volkov}}
\email{volkov@theor.jinr.ru}
\affiliation{Bogoliubov Laboratory of Theoretical Physics, \\
Joint Institute for Nuclear Research, 141980 Dubna, Russia}

\begin{abstract}
We consider a nonlocal version of the Nambu and Jona-Lasinio model. The
nonlocality is contained in the quark-antiquark bilinears of the four-quark
vertices as a form factor of the Gaussian type. The model has three parameters
which can be fixed in favour of the values of the pion mass, the pion decay
constant $f_\pi$, and the current quark mass. The $\pi\pi$ scattering amplitude
is obtained by calculating the quark box and the $\sigma$-pole diagrams, where
$\sigma$ is the scalar isoscalar meson. It is shown that this amplitude
satisfies the well-known Weinberg relation. We obtain the $s,p,d$ wave
scattering lengths in all isotopic channels and the $s$ wave slope parameters.
The results are in satisfactory agreement with both phenomenological data and
the basic requirements of low-energy theorems, thus supporting to the form
factor used.
\end{abstract}

\maketitle

\section{Introduction}

The theoretical study of $\pi\pi$ scattering has a long-standing history. One
of the pioneering works which has certainly boosted the development of chiral
symmetric models was due to Weinberg \cite{Weinberg:1966kf}. In that paper the
methods of current algebra \cite{CA:1973,Alfaro:1973} were successfully used to
compute the low-energy $\pi\pi$ scattering lengths. After that many Lagrangian
models of strong interactions of hadrons were intensively developed, using the
linear \cite{Gasiorowicz:1969kn} and nonlinear
\cite{Wess:1969,Callan:1969sn,VPBook} realizations of chiral symmetry. Various
low-energy properties of hadrons were described in the framework of these
models, for instance $\pi\pi$ scattering (see e.g.
\cite{Lehmann:1972kv,Pervushin:1974nm,VP:1975}).

The chiral perturbation theory has been developed on the basis of nonlinear
Lagrangians. This effective field theory describes the low-energy structure of
different amplitudes in terms of expansion in powers of energies, momenta and
current quark masses \cite{Weinberg:1978kz,Gasser:1983yg}. Chiral symmetry
determines the low energy behaviour of the $\pi\pi$ scattering amplitude to
within very small uncertainties \cite{Colangelo:2001,Colangelo:2001df}. For
such a theory one takes for granted that chiral symmetry is spontaneously
broken. There are theoretical arguments indicating that chromodynamics leads
indeed to the formation of a quark condensate \cite{Vafa:1984}. However, the
chiral perturbation theory cannot explain this phenomenon.

An analogy with superconductivity led Nambu and Jona-Lasinio (NJL) to the field
theoretical model of strong four-fermion interactions
\cite{Nambu:1961,Nambu:1961b} which provide for spontaneous chiral symmetry
breaking and the formation of a quark condensate. The local version of this
model was particularly well developed at the quark level
\cite{Eguchi:1976iz,Volkov:1982zx,Ebert:1982pk,Volkov:1984,Volkov:1986,Ebert:1985kz,Klimt:1989pm,Klevansky:1992qe}.
The spectrum, low-energy dynamics, the main strong and electromagnetic decays,
scattering and the internal characteristics of mesons have a reasonable
explanation within this model.

In this paper, we generalize the local approach to $\pi\pi$ scattering by
constructing the $\pi\pi$ scattering amplitude within the framework of a
nonlocal NJL model with linear realization of chiral symmetry. The scalar and
pseudoscalar degrees of freedom are considered. Let us note that one needs to
know higher powers of the $p^2$ expansion to describe scattering lengths and
slope parameters of $d$-waves and higher. Such expansions were first a problem
in the local approach: ad hoc form factors were inserted into the meson-quark
vertices \cite{VolkovOs:1984}, their structure not being derived from the
internal structure of the model, thus requiring justification
\cite{Volkov:1996rc}.

A coherent method which takes into account the whole contribution of quark one-loop
diagrams in the local NJL model has been formulated in
\cite{Bernard:1992mp,Bernard:1995hm}. In particular, the method was successfully used to
calculate $\pi\pi$ scattering lengths.

Recently, nonlocal versions of the NJL model have been actively developed
\cite{Andrianov:1993,Efimov2,Birse:1998,Celenza:1999cx,Anikin:2000rq,Radzhabov:2003hy,Scarpettini:2004,Ito:1991pv,Roberts:1993ks,Tandy:1997qf,Blaschke:2000gd,Faessler:2003yf,Blaschke:2005kx}.
These models are free from ultra-violet divergences, as well as there is hope
to confine the quarks because the constituent quark mass depends on the
momentum. The form factors of the meson-quark vertices are the direct
consequence of this nonlocality. Thus, it seems interesting to calculate again
$\pi\pi$ scattering lengths and parameters of effective ranges in the framework
of a nonlocal model. The relative simplicity of the process together with a
large amount of experimental information about $\pi\pi$ scattering serve us to
gain a considerable insight into the structure of nonlocal interactions from
such calculations.

The paper is organized as follows. In the next Sect. we consider the effective
$SU(2)\times SU(2)$ symmetric nonlocal four-quark interaction. The dynamical
mechanism of spontaneous chiral symmetry breaking is clarified after
bosonization. We show that the model satisfies Goldberger -- Treiman and
Gell-Mann, Oaks, Renner relations. We fix parameters of the model in this
section. In Sect. 3 we obtain the $\pi\pi$ scattering amplitude, check Weinberg
relation and calculate $s$, $p$, $d$ scattering lengths in all isotopic
channels and slope parameters of $s$-waves. The results are compared with
empirical data. In the last Sect. we discuss them.

\section{Nonlocal model of the NJL type}

The model is based on the effective $SU(2)\times SU(2)$ symmetric
quark interaction in the scalar and pseudoscalar sectors
\begin{equation}
   \mathcal{L}= \bar{q}(x)(i \hat{\partial}_x -m^0)q(x) + \frac{G}{2}
   \left( J_\sigma(x) J_\sigma(x) + J_\pi^a(x) J_\pi^a(x)
   \right),\label{lag}
\end{equation}
where $q, \bar{q}$ are quark and antiquark fields, $m$ are masses of current
quarks, $\tau_i$ are the Pauli matrices and $G$ is the four-quark coupling
constant. The nonlocal quark currents $J_I(x)$ are given by\footnote{The models
with such nonlocality have been considered previously in
\cite{Ito:1991pv,Roberts:1993ks,Tandy:1997qf,Blaschke:2000gd,Faessler:2003yf,Blaschke:2005kx}
}
\begin{equation}
   J_I(x) = \int d^4x_1 d^4x_2 \,
   \delta\left(x-\frac{x_1+x_2}{2}\right)f((x_1-x_2)^2)\,
   \bar{q}(x_1) \, \Gamma_I \, q(x_2) , \label{Jnl}
\end{equation}
where the function $f(x)$ is normalized as $f(0)=1$. The matrices
$\Gamma_I$ in eq.(\ref{lag}) are defined as follows
$\Gamma_\sigma=\mathbf{1}$, $\Gamma_\pi^a=i \gamma^5 \tau^a$.

One has for this Lagrangian after bosonization
\begin{equation}
 \mathcal{L} =
  \bar{q}(x)(i \hat{\partial}_x -m_c)q(x)  - \frac{1}{2 G}\left( \pi^a(x)^2
  +\tilde{\sigma}(x)^2\right)+J_\sigma(x)\tilde{\sigma}(x)+\pi^a(x)J_\pi^a(x),
\end{equation}
where $\tilde{\sigma}$ and $\pi^a$ are the scalar and pseudoscalar fields. The
scalar field  $\tilde{\sigma}$ has a nonzero vacuum expectation value
$\langle\tilde{\sigma}\rangle_0=\sigma_0\neq0$. To arrive at the physical
scalar field with zero expectation value, it is necessary to shift the field:
$\tilde{\sigma}=\sigma+\sigma_0$.

Varying the action with respect to $\sigma$
\begin{equation}
   \left\langle \frac{\delta \mathcal{S}}{\delta
   \sigma}\right\rangle_0 = 0,
\end{equation}
one obtains the equation for the dynamical quark mass, the ``gap'' equation
\begin{equation}
   m(p^2)=m_c+i G \frac{N_f  N_c }{(2 \pi)^4}f(p^2) \int \frac{d^4
   k}{(2\pi)^4} f(k^2) \mathrm{Tr}\left[S(k)\right]\label{gap},
\end{equation}
where $N_f$ and $N_c$ are the numbers of quark flavours and colours,
respectively. Note that due to the form factors the quark loop integral is
finite. This equation can also be written as
\begin{equation}
   m(p^2)=m_c-\sigma_0 f(p^2)=m_c+(m_q-m_c) f(p^2),\label{gap1}
\end{equation}
where $m_q$ is a dimensionfull parameter which plays the role of the
constituent quark mass. As a result the quark mass becomes $p^2$ dependent. The
quark propagator $S(p)$ has the form
\begin{equation}
   S(p) = \frac{1}{\hat p - m(p^2)}. \label{QP}
\end{equation}

The propagators of the meson fields are
\begin{equation}
   D_{\sigma,\pi}(p^2)=\frac{1}{-G^{-1}+J_{\sigma,\pi}(p^2)}=
   \frac{g^{2}_{\sigma,\pi}(p^2)}{p^2-M_{\sigma,\pi}^2}
\label{mesonprop},
\end{equation}
where $M_{\sigma,\pi}$ are meson masses, $g_{\sigma,\pi}(p^2)$ are the
functions describing the renormalization of mesonic fields, and
$J_{\sigma,\pi}(p^2)$ are the contributions of the quark one-loop diagrams
\begin{equation}
   J_{\sigma,\pi}(p^2)=i \frac{N_f N_c}{(2\pi)^4} \int d^4 k
   f^2(k^2)\, \mathrm{Tr}\left[ \mathrm{S}(k_-)\Gamma_{\sigma,\pi}
   \mathrm{S}(k_+) \Gamma_{\sigma,\pi} \right],
\label{poloper}
\end{equation}
where $k_+=k+p/2$, $k_-=k-p/2$. The meson masses $M_{\sigma,\pi}$ are the poles
of the propagators
\begin{equation}
   J_{\sigma,\pi} (M_{\sigma,\pi}^2) = G^{-1}, \label{pii}
\end{equation}
and the couplings $g_{\sigma,\pi}(M_{\sigma,\pi}^2)$ are determined on
the mass-shell of the mesonic states by the formulae
\begin{equation}
   g_{\sigma,\pi}^{-2}(M_{\sigma,\pi}^2) = \left.\frac{d
   J_{\sigma,\pi} (p^2)}{d
   p^2}\right|_{p^2=M_{\sigma,\pi}^2}.
\label{gm}
\end{equation}

The amplitude of the pion decay $\pi\rightarrow\mu\nu$ is
\begin{equation}
   A_{(\pi \rightarrow \mu \nu)}^\mu(p)=i p^\mu f_\pi,
\end{equation}
where $f_\pi$ is the coupling of the weak pion decay\footnote{Detailed
description of insertion of electroweak fields into nonlocal Lagrangian is
given in appendix.}. The Goldberger -- Treiman relation is held in the chiral
limit $f_\pi={m_q}/{g_\pi(0)}$ and $f_\pi$ is given by the formula of Pagels
and Stokar \cite{Pagels:1979hd}
\begin{equation}
   f_\pi^2=\frac{N_c}{4\pi^2}\int \limits_0^\infty du \, u
   m(u)\frac{m(u)-\frac{1}{2} u m^{\prime}(u)}{(u+m^2(u))^2}.
\end{equation}
This expression is given in Euclidean domain $u=-p^2$.

The Gell-Mann -- Oaks -- Renner relation is fulfilled in the model
\begin{equation}
   M_\pi^{2}f_\pi^2 = -2\, m_c \langle \bar{q}q \rangle + O(m_c^2),
\end{equation}
where $\langle \bar{q}q \rangle$ is the quark condensate. This relation can be
obtained from the equation for the pion mass (\ref{pii}) with the help of the
gap equation (\ref{gap}) and expanding in powers of $M_\pi^2$ and $m_c$. The
form factor $f(u)$ has been chosen in the form of a Gaussian exponent
$f(u)=\exp(-u/\Lambda^2)$.

The model has three parameters $m_c$, $\Lambda$, $G$. After using the values of
pion mass $M_\pi=140$ MeV and the weak pion decay constant $f_\pi=92$ MeV as
input parameters we obtain instead of three arbitrary parameters only one
$m_c$. This parameter can be chosen in the interval $4.05 - 7.425$ MeV at the
typical hadron scale 1 GeV \cite{Eidelman:2004wy}.
Additional restriction can be obtained, e.g., from the $d$-wave scattering
length $a_2^2$. As a result, the parameters of the model are $m_c=4.8$ MeV,
$\Lambda=880$ MeV and $G=22.87$ GeV$^{-2}$. Thus, one can find that $m_q=345$
MeV and $\langle \bar{q}q \rangle=(-255 \mathrm{MeV})^3$. For the mass and the
width of scalar meson we obtain $M_\sigma=570$ MeV, $\Gamma_\sigma=220$ MeV.

\section{$\pi\pi$ scattering}

\begin{figure*}[t!]
\setcaptionmargin{5mm}
\onelinecaptionstrue
\resizebox{0.7\textwidth}{!}{\includegraphics{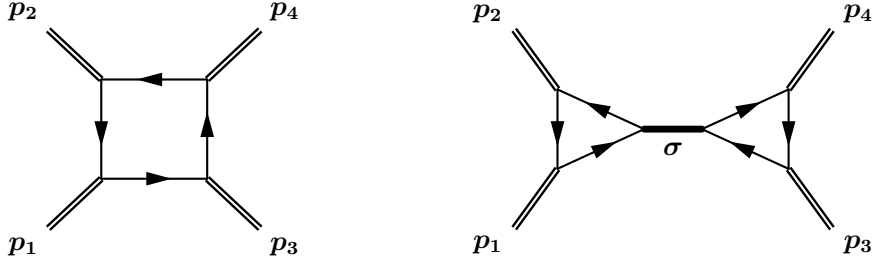}}
\captionstyle{normal} \caption{Diagrams contributing to
the amplitude of $\pi\pi$ scattering.}
\label{pi-pi}
\end{figure*}

The amplitude of the process is described by two main diagrams plotted on
Fig.\ref{pi-pi}.

Using the notation of Chew and Mandelstam one can write the matrix
element of $\pi\pi$ scattering
\begin{equation}
   \langle i_1 i_2| \mathcal{A}| i_3 i_4\rangle =
   \delta_{i_1 i_2}\delta_{i_3 i_4}A(s,t,u) +
   \delta_{i_1 i_3}\delta_{i_2 i_4}A(t,s,u) +\delta_{i_1
   i_4}\delta_{i_2 i_3}A(u,t,s) %\right\}
\end{equation}
in terms of three kinematical invariants
\begin{equation}
   s=(p_1+p_2)^2,\quad t=(p_1-p_3)^2,\quad u=(p_1-p_4)^2.
\end{equation}
For the amplitudes with the definite isospin values $I=0,1,2$ one has
\begin{eqnarray}
   T^0&\! =\!&3 A(s,t,u) + A(t,s,u)+A(u,t,s)\nonumber\\
   T^1&\! =\!&A(t,s,u)-A(u,t,s)\\
   T^2&\! =\!&A(t,s,u)+A(u,t,s).
\nonumber
\end{eqnarray}

The  $\sigma$-pole diagram contains the $\sigma\pi\pi$ vertex which has the
following structure:
\begin{eqnarray}
   A_{\sigma\to\pi\pi}&\! =\!& 2g_{\sigma}(p^2)g_{\pi}(p_1^2)g_{\pi}
                           (p_2^2)g_{\sigma\pi\pi}(p^2,p_1^2,p_2^2)
                           \nonumber\\
   g_{\sigma\pi\pi}(p^2,p_1^2,p_2^2)&\! =\!&i \frac{N_f N_c}{(2\pi)^4}
   \int d^4 k f\left(\left(k+\frac{p_1}{2}\right)^2\right)f
   \left(\left(k-\frac{p_2}{2}\right)^2\right)f\left(\left(k+
   \frac{p_1-p_2}{2} \right)^2\right)
   \nonumber\\
   &&\times\mathrm{Tr}\left[ \mathrm{S}(k+p_1)\Gamma_{\sigma}
   \mathrm{S}(k-p_2) \Gamma_{\pi} \mathrm{S}(k) \Gamma_{\pi}
   \right],\nonumber
\end{eqnarray}

The box diagram contributes as
\begin{eqnarray}
   A_{box}(s,t,u) &\! =\!& 2g_\pi^4(M_\pi^2)\left(J_4(s,t,u)+J_4(s,u,t)
                       -J_4(u,t,s)\right) \nonumber\\
   J_4(s,t,u)&\! =\!& i\frac{N_f N_c}{(2\pi)^4} \int d^4 k
   f\left(\left(k+\frac{p_1}{2}\right)^2\right)f
   \left(\left(k-\frac{p_2}{2}\right)^2\right)f\left(
   \left(k+\frac{p_1-p_3}{2}\right)^2\right) \nonumber\\
   &&\times f\left(\left(k+\frac{p_1-p_2-p_3}{2}\right)^2
   \right)\mathrm{Tr}\left[
   \mathrm{S}(k)\Gamma_{\pi} \mathrm{S}(k-p_2)\Gamma_{\pi}
   \mathrm{S}(k+p_1-p_3) \Gamma_{\pi} \mathrm{S}(k+p_1) \Gamma_{\pi}
   \right], \nonumber
\end{eqnarray}

As a result, the whole $\pi \pi$ amplitude can be written as
\begin{equation}
   A(s,t,u)=
   A_{box}(s,t,u) + 4g_\pi^4(M_\pi^2)
   \frac{g_{\sigma\pi\pi}^2(s,M_\pi^2,M_\pi^2)}{G^{-1}-J_{\sigma}(s)}.
\label{Afull}
\end{equation}

At the lowest order of the momentum and $m_c$ expansions one obtains the
Weinberg result
\begin{equation}
A(s,t,u)=\frac{s-{M}_\pi^2}{{f}_\pi^2}+O(p^4)
\end{equation}

Let us recall that in the local NJL model the Weinberg formula can be obtained on the
basis of the box and the $\sigma$-pole diagrams, if one picks up only the divergent terms
of the quark one-loop diagrams \cite{VolkovOs:1984,Volkov:2002bv}
\begin{equation}
   A(s,t,u)=-4{g^2} + \frac{16m^2g^2}{M^2_{\sigma}-s} =
   \frac{s-M^2_\pi}{f^2_\pi}+ \ldots , \label{Api}
\end{equation}
where $g=g_\pi=g_\sigma$, $m$ is the constituent quark mass, and the sigma meson mass
takes its model value $M_{\sigma}^2=M_{\pi}^2+4m^2$.

Scattering lengths $a^I_l$ and slope parameters $b^I_l$ are the first terms in the
expansion of partial-wave amplitudes for orbital angular momentum $l$ and isospin $I$
\begin{equation}
   \lim\limits_{k^2 \to 0} \frac{1}{64 \pi M_\pi
   k^{2l}}\int\limits_{-1}^{1}dx P_l(x)
   T^{I}(s,t,u)=a^I_{l}+k^2b^I_{l}+O(k^4).
\label{sc-ln}
\end{equation}
Here $s=4(M_\pi^2+k^2),\, t=-2k^2(1-x),\, u=-2k^2(1+x)$; $k$ is the c.m.
three-momentum, and $x=\cos\theta$, where $\theta$ is a scattering angle
between pions in the c.m. frame and $P_l(x)$ are the Legendre polynomials.

To obtain scattering lengths we expand the integrand of eq.(\ref{sc-ln}) in
powers of $k^2$. The subsequent integration over $x$ lead to the expressions
for the scattering lengths and slope parameters that are given in appendix.

The obtained values for the scattering lengths of the $s,p$ and  $d$ waves in
all isotopic channels as well as the parameters of the effective ranges for the
$s$ waves are given in Table.% \ref{t_sc_ln}.
To compare our results, we also present there characteristics obtained in the
framework of the local NJL model \cite{Bernard:1992mp}. The empirical data are
taken from the papers \cite{Adeva:2005pg} (the life-time of the $\pi^+ \pi^-$
mesonic atom), \cite{Batley:2005ax} (the $K\to 3\pi$ decay),
\cite{Pislak:2003sv} (the $K^\pm\to \pi^+ \pi^- e^\pm \nu_e (\bar{\nu}_e)$
decay), and \cite{Pelaez:2004vs} (the phenomenological analysis of data).

To check the result, we passed to the local limit in our formulae, by
regularizing the quark one-loop diagrams in the standard way and taking
$\Lambda\to\infty$ in the nonlocal form factors. After this the results of
known calculations in the local model are reproduced \cite{Bernard:1992mp}.

\begin{table}
 \setcaptionmargin{0mm}
 \onelinecaptionstrue
 \captionstyle{flushleft}
 \caption{Scattering lengths and slope parameters for the $\pi\pi$ system.
  Colums LNJL, PY and EXP correspond to the results obtained in the
  local NJL model \cite{Bernard:1992mp}, results of the phenomenological
  analysis of experimental data \cite{Pelaez:2004vs}, and known data
  extracted from the life-time of $\pi^+ \pi^-$ mesonic atom
  \cite{Adeva:2005pg}, from the $K\to 3\pi$ decay \cite{Batley:2005ax}
  and $K^\pm\to \pi^+ \pi^- e^\pm \nu_e (\bar{\nu}_e)$
  data \cite{Pislak:2003sv}.} \label{t_sc_ln}
\begin{tabular}{|c|c|c|c|c|c|c|c|c|c|}
\hline
                     &This work&LNJL&PY               & EXP   \\
\hline
$a_0^0 M_\pi$              &$0.20$  &$0.190$  &$0.230\pm0.015$  &\begin{minipage}{4cm}$0.216\pm0.017$\cite{Pislak:2003sv}  \end{minipage} \\
$a_0^2 M_\pi$              &$-0.049$ &$-0.044$ &$-0.048\pm0.0046$&\begin{minipage}{4cm}\vspace{0.3cm}$-0.0454\pm0.0049$  \cite{Pislak:2003sv}\\\vspace{-0.2cm}  $-0.041\pm0.036$\cite{Batley:2005ax}  \vspace{0.3cm}\end{minipage} \\
$(a_0^0-a_0^2)M_\pi$        &$0.25$  &$0.234$ &$0.277\pm0.014$  &\begin{minipage}{4cm}\vspace{0.3cm} $0.264^{+0.033}_{-0.020}$  \cite{Adeva:2005pg} \\\vspace{-0.2cm} $0.268\pm0.027$\cite{Batley:2005ax} \vspace{0.3cm}\end{minipage}   \\
$a_1^1\times 10^{-3}M_\pi^3$&$37$     &$34$     &$38.4\pm0.8$     &              \\
$a_2^0\times 10^{-4}M_\pi^5$&$16.3$   &$16.7$   &$18.7\pm0.41$    &                \\
$a_2^2\times 10^{-4}M_\pi^5$&$2.8$    &$3.2$    &$2.78\pm0.37$    &                \\
$b_0^0 M_\pi^3$              &$0.26$  &$0.27$  &$0.312\pm0.014$  &                \\
$b_0^2 M_\pi^3$              &$-0.085$ &$-0.079$ &$-0.090\pm0.006$ &                \\
\hline
\end{tabular}
\end{table}

\section{Conclusions}

The present calculations show that the main low energy theorems of
pion physics are fulfilled within the nonlocal NJL model extension
considered. We have checked the Goldberger -- Treiman relation, the
Gell-Mann, Oaks, Renner formula and the famous Weinberg result for
the low energy $\pi\pi$ scattering amplitude.

The Goldberger -- Treiman relation is an exact result of the theory in
the Goldstone limit and reflects the chiral asymmetry of the ground
state. In the NJL model as well as in the linear sigma model, the
Goldberger -- Treiman relation can be understood as
$\langle\sigma_0\rangle = f_\pi$, where $\langle\sigma_0\rangle$
represents the point at which the effective potential has a minimum.

The Gell-Mann, Oaks, Renner formula in the NJL model relates the quark condensate with
the gap equation, thus reflecting the dynamical mechanism for the spontaneous chiral
symmetry breaking.

The $\pi\pi$ scattering amplitude has been obtained in the Hartree-Fock
approximation of the nonlocal NJL model and was used to calculate the threshold
characteristics. The results for the scattering lengths in $s,p$ and $d$ waves
and for the $s$-wave slope parameters are in satisfactory agreement with the
empirical data.

The main contribution to the values of the $s,p$ scattering lengths is given by the $p^2$
terms of the amplitudes. By virtue of the Weinberg relation, they are less sensitive to
the structure of the form factor. The $d$-waves are mainly determined by the $p^4$ order
terms of the amplitude and the sufficient predictions for the $d$-waves are a non-trivial
result of our calculations. In terms of box and $\sigma$-pole diagrams, they are additive
for the $a_2^0$ scattering length and have opposite signs in the $a_2^2$ case. The value
of the $a_2^2$ scattering lengths is strongly affected by the magnitude of the $\sigma$
meson mass predicted by the model.

Returning to the problem of fixing model parameters, note that the best
agreement with experimental $a_2^2$ is achieved for $m_c=4.8$
MeV($a_2^2=2.8\times 10^{-4} M_\pi^{-5}$). For the current quark mass equal to
$4.4$ and $5.2$ MeV this scattering length equals $3.5\times 10^{-4}
M_\pi^{-5}$ and $2\times 10^{-4} M_\pi^{-5}$, respectively.

In the future we plan to extend the nonlocal model considered by including the
vector and axial-vector mesons and taking into account the $\pi-a_1$
transitions. It also will be interesting to consider the $U(3)\times U(3)$
chiral symmetry group and study the internal structure and interactions of the
pseudoscalar, scalar, vector, and axial-vector nonets. Of special interest will
be processes in which the $p^2$ dependence of the amplitudes will be relevant.

Moreover, it is interesting to investigate the $\pi\pi$ scattering process in
the hot and dense medium in the framework of a nonlocal model. When chiral
symmetry is restored, the $\sigma$ meson mass tends to the mass of pion
\cite{Blaschke:2005kx,Volkov:2002bv,Ebert:1992ag}. This can lead to a
noticeable increase in the $\pi\pi$ scattering amplitude near the phase
transition of hadron matter into quark-gluon plasma. Such investigations in the
framework of local models are performed in \cite{Jido:2000bw}.

The authors thank A. E. Dorokhov for useful discussions.

A.E.R. and M.K.V. acknowledge the support of the Russian Foundation for Basic
Research, under contract 05-02-16699.

A. A. Osipov acknowledges support from grants provided by Fundac\~ao para a
Ci\^encia e a Tecnologia, POCTI/FNU/50336/2003 and POCI/FP/63412/2005. This
research is part of the EU integrated infrastructure initiative Hadron Physics
project under contract No.RII3-CT-2004-506078.

\section*{Appendix}
\subsection*{Weak pion decay constant}

To determine the weak pion decay constant one inserts external sources
associated with the electroweak fields. For this purpose it is necessary to
take into account the contributions induced by the nonlocal interaction. We use
the standard procedure: the external fields are introduced via the Schwinger
phase factor which delocalizes quark fields
\begin{equation}
   q(y)\to Q(x,y)= \mathcal{P}\mathrm{exp}\left\{i\int \limits_x^y
   dz^\mu [V^a_\mu(z)+ A^a_\mu(z)\gamma_5]T^a\right\} q(y).
\end{equation}
Here $\mathcal{P}$ is the Dyson ordering along the path, $V^a_\mu(z)$ and $A^a_\mu(z)$
are the external vector and axial-vector gauge fields which can be identified with
electroweak fields, $T^a$ stands for the generators of the flavour group. Since the
derivative of the path integral of any function $F_{\mu }(z)$ over the path does not
depend on the trajectory along which one calculates the integral \cite{Mandelstam:1962mi}
\begin{equation}
\frac{\partial }{\partial y^{\mu }}\int\limits_{x}^{y}dz^{\nu }\ F_{\nu
}(z)=F_{\mu }(y),\qquad \delta ^{(4)}\left( x-y\right)
\int\limits_{x}^{y}dz^{\nu }\ F_{\nu }(z)=0. \label{intrules}
\end{equation}
As a result, non-minimal interactions, which could be induced by the kinetic
term of quark fields, are absent. The gauge invariant procedure used here leads
to additional vertices with any number of external fields which have their
origin in the nonlocal nature of the interaction.

\begin{figure*}[t!]
\setcaptionmargin{5mm}
\onelinecaptionstrue
\resizebox{0.7\textwidth}{!}{\includegraphics{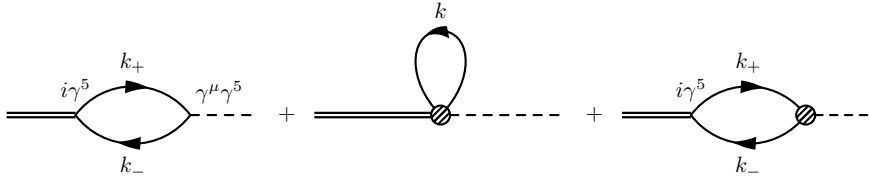}} \captionstyle{normal}
\caption{Diagrams corresponding to the weak decay of the
         pion $\pi \rightarrow \mu \nu$.}%
\label{fpi}%
\end{figure*}

Weak pion decay is described by three diagrams (see fig.\ref{fpi}).
Note that in the chiral limit the contributions of nonlocal diagrams
are exactly cancelled.

\subsection*{Scattering lengths and slope parameters}

\begin{eqnarray}
a_0^2  &=&\frac{1}{32 \pi M_\pi}\biggl( A(0,4M_\pi^2,0)+A(0,0,4M_\pi^2) \biggr)\nonumber\\
(a_0^0-a_0^2) &=&\frac{3}{32 \pi M_\pi} A(4M_\pi^2,0,0) \nonumber\\
a_1^1  &=& \frac{1}{48 \pi M_\pi}
\biggl(A^{(s)}(0,4M_\pi^2,0)+A^{(s)}(0,0,4M_\pi^2)-A^{(t)}(0,0,4M_\pi^2)-A^{(u)}(0,4M_\pi^2,0)\biggr)\nonumber\\
a_2^2 &=& \frac{1}{120 \pi M_\pi}\biggl(A^{(u,u)}(0,4M_\pi^2,0)+A^{(t,t)}(0,0,4M_\pi^2)-2A^{(s,u)}(0,4M_\pi^2,0)- \biggr.\nonumber\\
&&\biggl.\quad\quad\quad\quad-2A^{(s,t)}(0,0,4M_\pi^2)+A^{(s,s)}(0,0,4M_\pi^2)+A^{(s,s)}(0,4M_\pi^2,0)\biggr) \nonumber\\
(a_2^0-a_2^2)  &=& \frac{1}{40 \pi M_\pi} \biggl(A^{(u,u)}(4M_\pi^2,0,0)-2A^{(t,u)}(4M_\pi^2,0,0)+A^{(t,t)}(4M_\pi^2,0,0)\biggr)\nonumber%\\
\end{eqnarray}
\begin{eqnarray}
b_0^2 &=& \frac{1}{16 \pi M_\pi}\biggl(2A^{(u)}(0,0,4M_\pi^2)+2A^{(t)}(0,4M_\pi^2,0)-A^{(u)}(0,4M_\pi^2,0)\biggr.\nonumber\\
&&\biggl.\quad\quad\quad-A^{(t)}(0,0,4M_\pi^2)-A^{(s)}(0,4M_\pi^2,0)-A^{(s)}(0,0,4M_\pi^2)\biggr)\nonumber\\
(b_0^0-b_0^2)  &=&\frac{3}{16 \pi M_\pi}\biggl(2
A^{(s)}(4M_\pi^2,0,0)-A^{(t)}(4M_\pi^2,0,0)-A^{(u)}(4M_\pi^2,0,0)\biggr)\nonumber,
\end{eqnarray}
where
\begin{eqnarray}
 A^{(x)}(a,b,c)=\frac{\partial}{\partial x}A(s,t,u)|_{s=a,t=b,u=c}\,,\quad
 A^{(x,y)}(a,b,c)=\frac{\partial^2}{\partial x \partial
 y}A(s,t,u)|_{s=a,t=b,u=c}\,.\nonumber
\end{eqnarray}

\end{document}